\documentclass{article} 
\usepackage{iclr2026_conference,times}

\usepackage[most]{tcolorbox}
\usepackage{xcolor}
\definecolor{abstractbg}{RGB}{235, 245, 255} 
\definecolor{metafg}{HTML}{1C2B33}

\tcbset{
  enhanced,
  colback=abstractbg,      
  colframe=metafg,     
  boxrule=0pt,             
  frame hidden,            
  arc=8pt,                 
  left=0pt,right=0pt,top=6pt,bottom=6pt,
  before skip=8pt, after skip=12pt,
}

\newenvironment{boxedabstract}
{\begin{tcolorbox}\begin{abstract}}
{\end{abstract}\end{tcolorbox}}

\usepackage{amsmath,amsfonts,bm}

\def\eqref#1{equation~\ref{#1}}

\def\1{\bm{1}}

\DeclareMathAlphabet{\mathsfit}{\encodingdefault}{\sfdefault}{m}{sl}
\SetMathAlphabet{\mathsfit}{bold}{\encodingdefault}{\sfdefault}{bx}{n}

\usepackage{multirow}
\usepackage[utf8]{inputenc} 
\usepackage{url}            
\usepackage{tabularx} 
\usepackage{array}
\renewcommand{\arraystretch}{1.1}
\usepackage[T1]{fontenc}    
\usepackage{booktabs}       
\usepackage{amsfonts}       
\usepackage{nicefrac}       
\usepackage{microtype}      
\usepackage{xcolor}         

\usepackage{multirow}
\usepackage[table,xcdraw]{xcolor}

\definecolor{metablue}{HTML}{0064E0}

\usepackage{caption}
\usepackage{subcaption}
\usepackage{natbib}
\usepackage{graphicx}
\usepackage{enumitem}
\usepackage{makecell}
\usepackage{amsmath}
\usepackage{amsthm}
\usepackage{thmtools}
\usepackage{thm-restate}
\usepackage{algorithm}
\usepackage{algorithmic}
\usepackage{bbm}
\usepackage{bm}
\usepackage{bbding} 

\usepackage{graphicx}
\usepackage{wrapfig}
\usepackage{multirow}
\usepackage{tcolorbox}
\usepackage{adjustbox} 
\usepackage{xcolor}
\usepackage{float}
\usepackage[normalem]{ulem}
\useunder{\uline}{\ul}{}

\RequirePackage{hyperref}
\RequirePackage[noabbrev,nameinlink]{cleveref}
\hypersetup{
  colorlinks,
  linkcolor=metablue,
  citecolor=metablue,
  urlcolor=metablue
}

\newcommand{\email}[1]{\href{mailto:#1}{\texttt{#1}}}

\newcommand{\slh}[1]{{\color{black}{#1}}}
\usepackage{fontawesome} 

\usepackage{graphicx}
\usepackage{fancyhdr}

\newcommand{\HeaderIconHeight}{12.3pt}

\renewcommand{\headrulewidth}{1pt}
\newcommand{\HeadRuleRaise}{0pt} 
\makeatletter
\renewcommand{\headrule}{%
  \vspace*{-\HeadRuleRaise}
  \hrule height \headrulewidth \@width\headwidth
  \vspace*{-\headrulewidth}
}
\makeatother

\fancypagestyle{plain}{%
  \fancyhf{}
  \fancyhead[LO,RE]{%
    \raisebox{0pt}[0pt][0pt]{\includegraphics[height=\HeaderIconHeight]{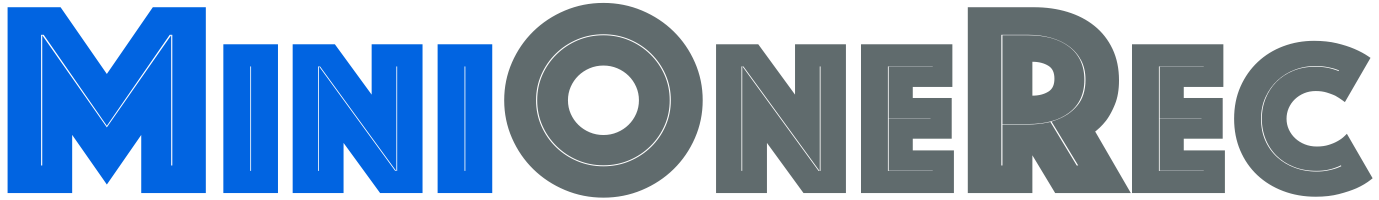}}%
  }

  \renewcommand{\headrulewidth}{0.4pt}%
  \makeatletter
  \renewcommand{\headrule}{%
    \vspace*{-\HeadRuleRaise}%
    \hrule height \headrulewidth \@width\headwidth
    \vspace*{-\headrulewidth}%
  }
  \makeatother

  \fancyfoot[C]{\thepage}
}

\title{MiniOneRec: An Open-Source Framework for \\Scaling Generative Recommendation}

\author{\textbf{Xiaoyu Kong}$^{\star}$,\quad\textbf{Leheng Sheng}$^{\circ}$,\quad\textbf{Junfei Tan}$^{\star}$,\quad\textbf{Yuxin Chen}$^{\circ}$,\\
\textbf{Jiancan Wu}$^{\star}$,\quad\textbf{An Zhang}$^{\star}$,\quad\textbf{Xiang Wang}$^{\star}$,\quad\textbf{Xiangnan He}$^{\star}$\\
  $^\star$University of Science and Technology of China~~
  $^\circ$National University of Singapore\\
  \email{\{kongxy,sober\_clever\}@mail.ustc.edu.com},\\
  \email{\{leheng.sheng,yuxin.chen\}@u.nus.edu},\\
  \email{\{wujcan,an.zhang3.14,xiangwang1223,xiangnanhe\}@gmail.com}
}

\iclrfinalcopy 
\begin{document}

\maketitle

\begin{boxedabstract}

The recent success of large language models (LLMs) has renewed interest in whether recommender systems can achieve similar scaling benefits. Conventional recommenders, dominated by massive embedding tables, tend to plateau as embedding dimensions grow. In contrast, the emerging generative paradigm replaces embeddings with compact Semantic ID (SID) sequences produced by autoregressive Transformers. Yet most industrial deployments remain proprietary, leaving two fundamental questions open: (1) Do the expected scaling laws hold on public benchmarks? (2) What is the minimal post-training recipe that enables competitive performance?

We present \textbf{MiniOneRec}, to the best of our knowledge, the first fully open-source generative recommendation framework, which provides an end-to-end workflow spanning SID construction, supervised fine-tuning, and recommendation-oriented reinforcement learning. We generate SIDs via a Residual Quantized VAE and post-train Qwen backbones ranging from 0.5B to 7B parameters on the Amazon Review dataset. Our experiments reveal a consistent downward trend in both training and evaluation losses with increasing model size, validating the parameter efficiency of the generative approach. To further enhance performance, we propose a lightweight yet effective post-training pipeline that (1) enforces full-process SID alignment and (2) applies reinforcement learning with constrained decoding and hybrid rewards. Together, these techniques yield significant improvements in both ranking accuracy and candidate diversity.

\faGithub\ \url{https://github.com/AkaliKong/MiniOneRec}

\hspace{-0.3em} \raisebox{-0.2em}{\includegraphics[height=0.9em]{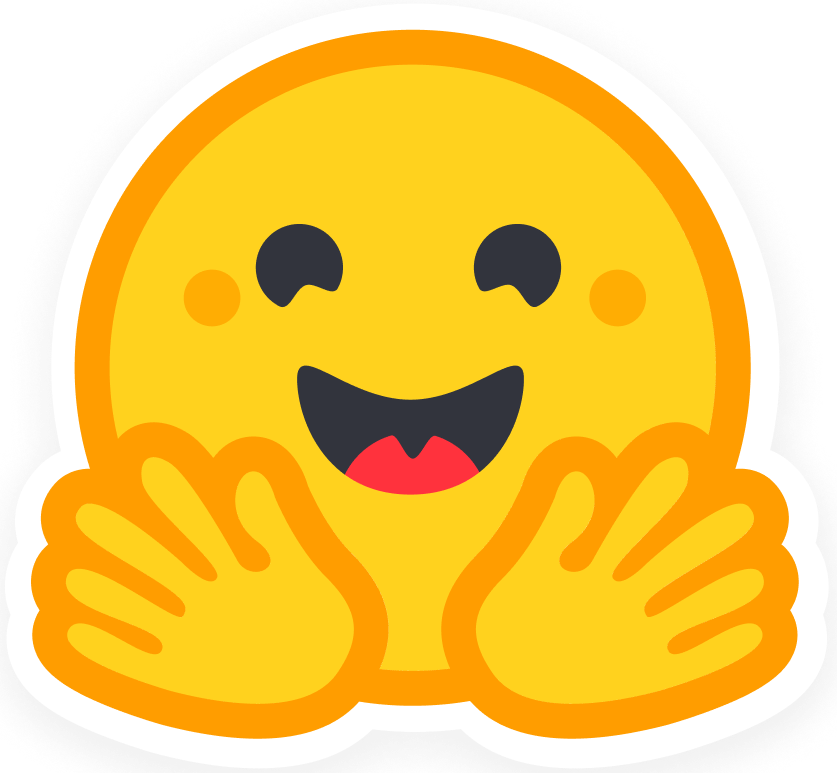}}
\url{https://huggingface.co/kkknight/MiniOneRec}

\end{boxedabstract}


\begin{center}
\begin{minipage}[t]{0.46\textwidth}
\centering
\includegraphics[width=\linewidth]{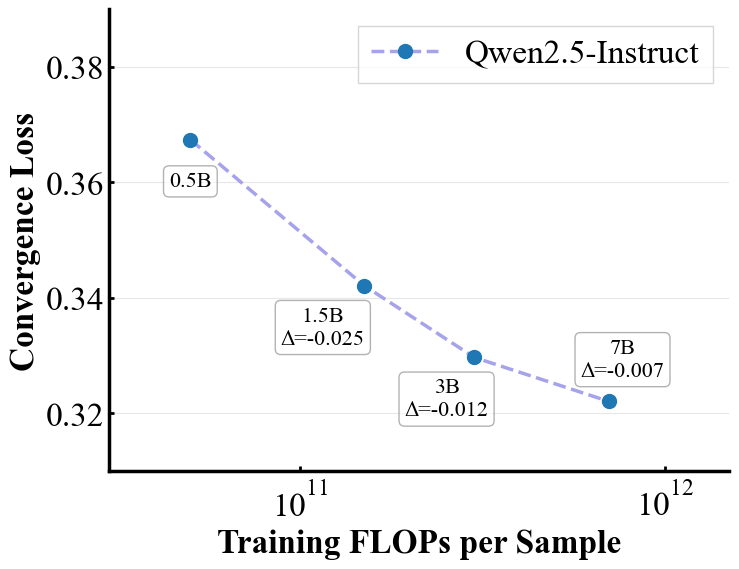}
\end{minipage}\hfill
\begin{minipage}[t]{0.46\textwidth}
\centering
\includegraphics[width=\linewidth]{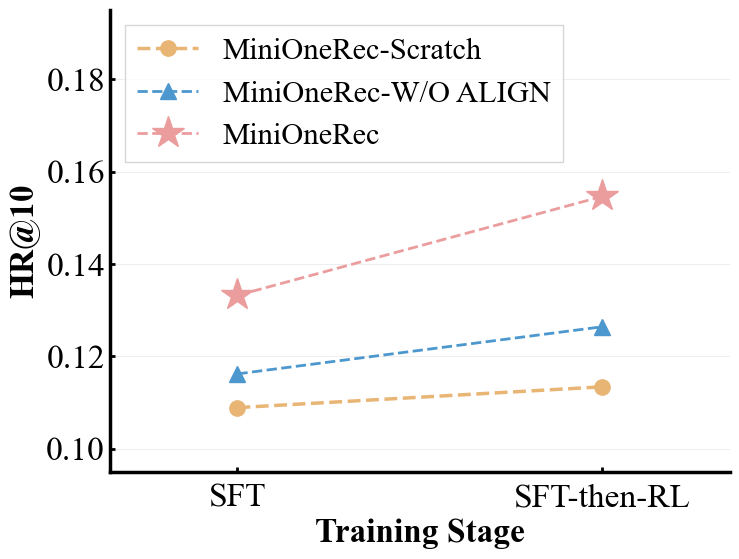}
\end{minipage}
\captionsetup{type=figure}
\captionof{figure}{Left: Scaling curves from 0.5B to 7B parameters. Right: Effect of world knowledge on model performance: MiniOneRec-W/O ALIGN uses pretrained LLM weights but omits SID–text alignment, while MiniOneRec-Scratch is trained from random initialization and omits alignment.}
\label{fig:scaling}
\end{center}

\clearpage
\tableofcontents
\clearpage

\section{Introduction}
The scaling behavior of recommendation models has recently attracted significant attention \citep{OneRec, mtgr, HSTU, Wukong, URM, RankMixer, HLLM, OnePiece}, driven largely by the success of large language models (LLMs) in demonstrating predictable performance gains with increased model size \citep{GPT3, gpt4, deepseek, qwen2.5, mistral-7b, llama3, Llama2, Deepseek-r1}.
However, traditional recommendation models have struggled to exhibit similar scaling laws \citep{SASRec, GRU4Rec, Caser, SGL, survey_seq}.
These systems typically allocate the majority of parameters to large embedding tables for storing user and item representations, while using only inner-product or shallow scoring networks for final predictions.
Such an embedding-heavy design leads to performance plateaus: even as embedding dimensions or table sizes increase, improvements diminish quickly beyond moderate scales \citep{DIN, DCN, DCNv2, DeepFM}.

Generative recommendation offers a fundamental paradigm shift.
By compressing items into sequences of discrete Semantic IDs (SIDs) through quantization techniques \citep{rqkmeans, rqvae}, it uses compact vocabularies and redirects the bulk of parameters toward deep autoregressive Transformers that generate SID sequences \citep{TIGER, LCRec, TOKENRec, OneRec, OnePiece}.
This design enables scaling behaviors more characteristic of language models, where increased depth and capacity translate to consistent performance gains \citep{OneRec}.

Recent industrial deployments have demonstrated the promise of this generative paradigm.
OneRec \citep{OneRec, onerec-tech-v1} achieves significant improvements on Kuaishou's platform with 400 million daily active users, while OneRec-V2 \citep{onerec-tech-v2} further advances through lazy decoder architecture and preference alignment. 
OnePiece \citep{OnePiece} integrates LLM-style context engineering and reasoning into both retrieval and ranking models of industrial cascaded pipelines.
However, these results rely on massive proprietary datasets and remain closed-source, leaving critical questions unanswered for the research community:
\begin{itemize}
    \item Do the scaling advantages of generative recommendation transfer to public datasets?
    \item What is the minimal post-training recipe needed to achieve strong performance?
\end{itemize}

In this work, we present \textbf{MiniOneRec}, the first fully open-source framework for generative recommendation, which provides an end-to-end workflow encompassing SID generation, model supervised fine-tuning (SFT), and recommendation-driven reinforcement learning (RL). The release includes complete source code, reproducible training pipelines, and publicly available model checkpoints.
We implement SID construction using residual quantized variational autoencoder (RQ-VAE) \citep{rqvae}, in line with prior work such as TIGER \citep{TIGER, OneRec}.
We post-train Qwen-based (backbone) generative models at multiple scales, ranging from 0.5B to 7B parameters, on public benchmarks from the Amazon Review Dataset.
Following OneRec's successful two-stage post-training pipeline, we adopt SFT on user-item interaction sequences, followed by group relative policy gradient (GRPO) \citep{Deepseek-r1, deepseekmath} for alignment.
The main contributions of our work are summarized as follows:
\begin{itemize}
    \item \textbf{Validating the Scaling Laws of Generative Recommendation}:
    We present the first systematic investigation of how generative recommendation models scale on public data, to the best of our knowledge. Specifically, we train four MiniOneRec variants, spanning 0.5\,B to 7\,B parameters, on the Amazon Review corpus~\citep{amazon} with identical data splits and hyper-parameter settings. Both the final training loss and the held-out evaluation loss consistently decrease as model size grows (see Figures \ref{fig:scaling} Left and \ref{fig:scaling_eval_loss}), highlighting the superior parameter efficiency of the generative paradigm.
    \item \textbf{Optimizing Post-training Strategies for Generative Recommendation}: 
    We design a lightweight yet comprehensive post-training pipeline that tightly couples full-process SID alignment with reinforced preference optimization. First, we verify the influence of world knowledge on the model’s generative recommendation performance (see Figure \ref{fig:scaling} Right) and augment the vocabulary with dedicated SID tokens and enforce auxiliary alignment objectives throughout the two-stage optimization process. Second, during the RL phase, we shape the generation process itself: invalid tokens are masked to guarantee that every step produces a legal item, beam search is employed for efficient exploration of diverse candidates, and the reward signal combines rule-based accuracy with a ranking-aware penalty that pushes the model away from hard negatives. As a result, MiniOneRec offers a concise yet effective recipe for bringing RL with value-based ranking into generative recommendation, simultaneously improving candidate diversity and ranking fidelity.
\end{itemize}

The remainder of this report is organized as follows:
Section \ref{bg} reviews background on generative recommendation and the application of LLMs with RL in recommender systems.
Section \ref{method} presents our proposed MiniOneRec, including task formulation, item tokenization, and post-training pipeline.
Section \ref{training_detail} describes the training details while Section \ref{results} provides comprehensive experimental results.
Section \ref{conclusion} concludes with discussions of future directions.

\section{Background and Related Work}
\label{bg}
\subsection{Generative Recommendation}

In the last few years, formulating recommendation as a sequence generation problem has become a vibrant research direction \citep{TIGER, GenRec-Direction}.  
Under this view, a recommender is trained to predict the tokens of the next item in an end-to-end manner, typically with a Transformer backbone \citep{Transformer}.  
Such a design removes the rigid \slh{multi-stage} “matching–ranking” pipeline in traditional retrieval-based recommender systems \citep{SASRec, GRU4Rec} and therefore pushes the performance upper bound.

Early explorations illustrate two key components: (1) turning an item into a discrete code (SIDs) and (2) letting the model produce the code.  
TIGER~\citep{TIGER} adopts RQ-VAE~\citep{rqvae} to map textual embeddings of titles and descriptions into SIDs.  
HSTU~\citep{HSTU} introduces a streaming architecture that is friendly to high-cardinality and non-stationary logs.  
LC-Rec~\citep{LCRec} aligns an LLM with the SIDs through multi-task learning so that the model can understand these symbols during generation.  

Subsequent work focuses on designing better codes.  
RecForest~\citep{recforest} clusters items via hierarchical $k$-means and uses the cluster indices as tokens.  
EAGER~\citep{eagar} and TokenRec~\citep{TOKENRec} fuse collaborative and semantic evidence directly into the tokenizer.  

At industrial scale, generative recommenders have started to replace heavy cascade systems.  
MTGR~\citep{mtgr} keeps the original DLRM features, adds user-level compression, and accelerates both training and inference.  
OneRec~\citep{onerec-tech-v2} reduces serving cost through a lazy decoder-only layout and stabilizes optimization with an improved RL algorithm.  
OnePiece~\citep{OnePiece} discovers that performing inference in the latent space can further improve generative recommendation performance.

\subsection{LLM and RL}

RL aims at maximizing cumulative reward through repeated interaction~\citep{RLsurvey}.  
When fine-tuning LLMs, RL with Human Feedback (RLHF) has become a standard recipe; PPO~\citep{PPO} is the most common optimizer but is memory-intensive for billion-scale parameters.  
To reduce cost, Direct Preference Optimization (DPO)~\citep{DPO} removes the separate value network and maximizes the log-likelihood gap between preferred and dispreferred outputs.  
S-DPO~\citep{sdpo} adapts this idea to recommendation by treating softmax-based negative sampling as implicit pairwise preference.  
Nevertheless, preference-based objectives are off-policy and may converge prematurely.

Light-weight on-line methods have therefore been proposed.  
GRPO~\citep{deepseekmath} normalizes rewards within a small batch of roll-outs and replaces the learned reward model by rule-based signals, achieving strong gains in maths and code generation.  

\section{Modeling}
\label{method}
\begin{figure}
    \centering
    \includegraphics[width=0.85\textwidth]{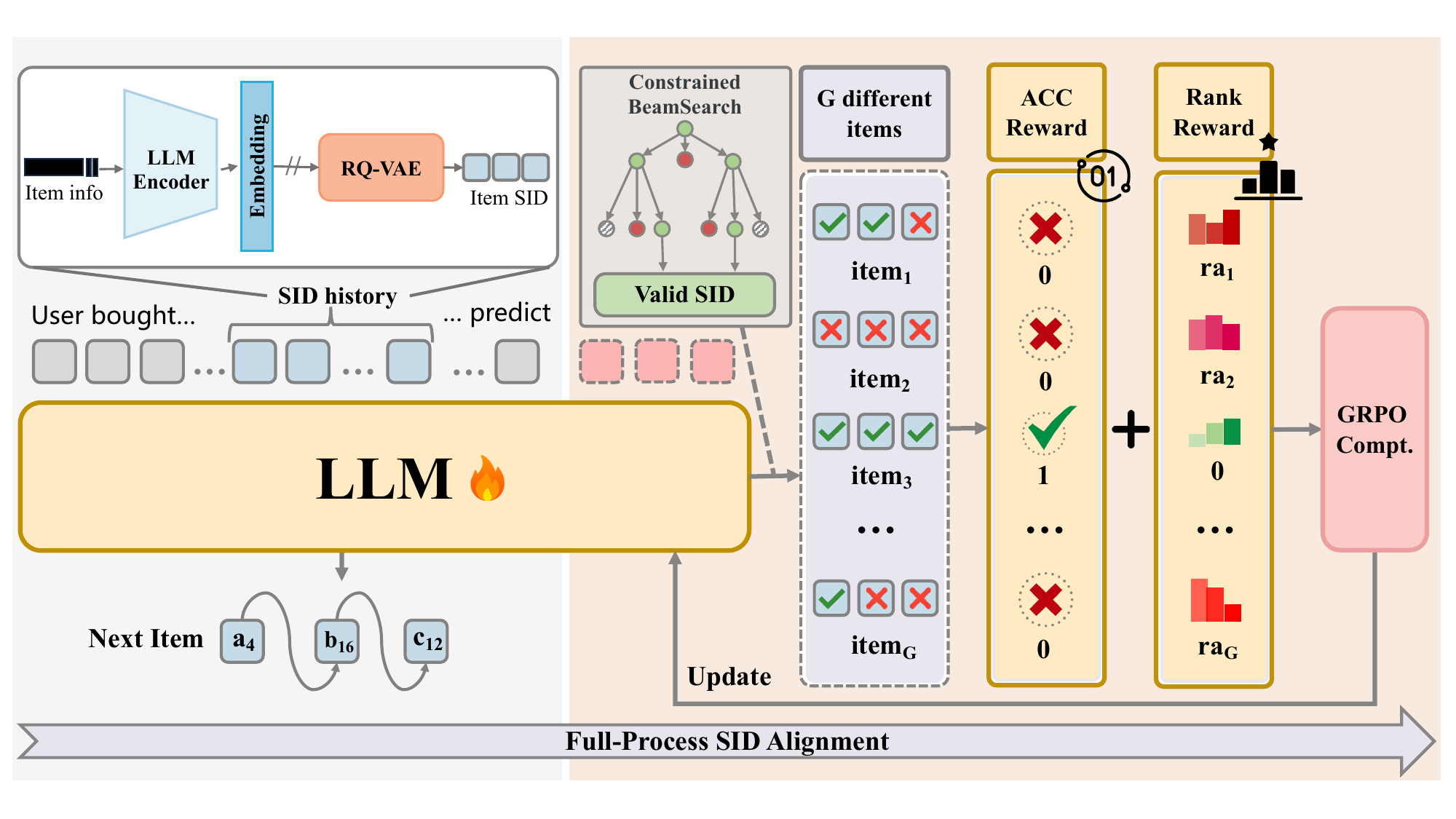}
    \vspace{-10pt}
    \caption{MiniOneRec framework. RQ-VAE builds the item SID codebook. We then perform SFT to warm up the LLM and obtain an initial alignment. In RL, beam search with constrained decoding, thereby the model sequentially produces a ranked list of distinct, valid SIDs. GRPO updates the policy, and SID alignment is enforced end-to-end. This alignment objective is preserved throughout both the SFT and RL stages, fostering deeper semantic understanding.}
    \label{fig:framework}
    \vspace{-10pt}
\end{figure}

In this section, we present the modeling strategies of \textbf{MiniOneRec}, as illustrated in Figure \ref{fig:framework}. 
MiniOneRec first converts the textual information into SIDs with the RQ-VAE tokenizer (Section \ref{sec:tokenization}). 
For better incorporating the huge world knowledge within LLMs \citep{LlaRA}, MiniOneRec further introduces the alignment with LLMs as one expansion of the original OneRec architecture (Section \ref{sec:alignment_llm}). 
During the training stage, MiniOneRec is optimized with next token prediction first, and then followed by reinforeced preference optimization (Section \ref{sec:optimization}).

\subsection{Task formulation}

We first formulate recommendation as a sequence–generation problem.  
For every user $u$, the items he or she has interacted with are sorted chronologically to form a sequence
$H_u=\bigl[i_1,i_2,\dots,i_T\bigr]$.  
Each item $i_t$ in the sequence is encoded by a three–level structural ID, $\bigl\{c_0^{i_t},\,c_1^{i_t},\,c_2^{i_t}\bigr\}$. Such structural IDs are typically called SIDs, which preserve hierarchical semantics through quantization techniques with semantic embeddings \citep{TIGER}.

A generative policy $\pi_\theta$, implemented as an autoregressive model with parameters $\theta$, reads the entire history $H_u$ and is trained to predict the next item $i^{+}$ that best matches the taste of user $u$ among all candidates in the catalog.  
During inference, the model recursively produces item tokens; we keep the $k$ most promising beams by the standard beam–search algorithm and return them as the recommendation list.  
Model performance is reported with the evaluation measures commonly adopted in generative recommendation.

\subsection{Item Tokenization} \label{sec:tokenization}

In SID-style generative recommenders, the first task is to convert each item into a sequence of discrete tokens. Following the practices of TIGER \citep{TIGER}, we employ RQ-VAE~\citep{rqvae} for this purpose. 
Concretely, the pipeline is:

\begin{enumerate}
    \item For every item $i$, we concatenate its title and textual description to form a single sentence;  
    \item This sentence is passed through a frozen text encoder \citep{AlphaRec}, producing a $d$-dimensional semantic vector $\mathbf{x}\in\mathbb{R}^{d}$;  
    \item Apply RQ-VAE to $\mathbf{x}$. At each level $l$ ($0\le l<L$) we have a separate codebook $\mathcal{C}_l=\{\mathbf{e}^{(l)}_{k}\}_{k=1}^{K}$, where $K$ is the codebook size. We set $L{=}3$ and $K{=}256$, so each item is represented by three bytes; this choice provides $2^{24}$ possible codes, which is sufficient for catalogs containing hundreds of millions of products while keeping the vocabulary small. The residual is initialized as $\mathbf{r}_{0}=\mathbf{x}$ and updated by
    \[
       c_l=\arg\min_{k}\left\|\mathbf{r}_l-\mathbf{e}^{(l)}_{k}\right\|_2,\qquad 
       \mathbf{r}_{l+1}=\mathbf{r}_l-\mathbf{e}^{(l)}_{c_{l}}.
    \]
    \item Collect the indices $(c_0,\dots,c_{L-1})$ as the discrete token sequence for item $i$; these indices constitute the item tokens consumed by the subsequent generative recommender.
\end{enumerate}

The quantized latent is reconstructed by
\[
    \mathbf{z}_{\text{q}}=\sum_{l=0}^{L-1}\mathbf{e}^{(l)}_{c_l},
\qquad
    \hat{\mathbf{x}}=D(\mathbf{z}_{\text{q}}),
\]
where $D(\cdot)$ is a decoder. The codebooks, the encoder and the decoder are trained jointly, while the text encoder remains frozen. The loss is the sum of a reconstruction term and an RQ regularizer:
\[
\mathcal{L}(\mathbf{x})=
\underbrace{\left\|\mathbf{x}-\hat{\mathbf{x}}\right\|_2^{2}}_{\mathcal{L}_{\text{RECO}}}
+\underbrace{\sum_{l=0}^{L-1}\!\bigl(
      \|\mathrm{sg}[\mathbf{r}_l]-\mathbf{e}^{(l)}_{c_l}\|_2^{2}
      +\beta\,\|\mathbf{r}_l-\mathrm{sg}[\mathbf{e}^{(l)}_{c_l}]\|_2^{2}\bigr)}_{\mathcal{L}_{\text{RQ}}},
\]
with $\mathrm{sg}[\cdot]$ the stop-gradient operator and $\beta$ a hyper-parameter controlling the commitment term.

To prevent codebook collapse, we follow the warm-start trick in prior work \citep{rqvae, LCRec} and initialize each codebook with $k$-means centroids computed on the first training batch.

\subsection{Alignment with LLMs} \label{sec:alignment_llm}
LLMs possess extensive understanding of the world and human behaviors 
\citep{AlphaRec}, which can serve as a supplement to the collaborative signals in recommender systems \citep{RLMRec}.
Existing work has shown that linking an LLM’s world knowledge to SID representations noticeably strengthens generative recommendation \citep{LlaRA, LCRec}. Therefore, rather than training on SIDs alone as in prior work \citep{TIGER, OneRec, mtgr}, we introduce several alignment objectives that tie the language space to SID signals. Two major groups of tasks are employed:  
\begin{itemize}
    \item \textbf{Recommendation Tasks}: The LLM receives a time-ordered history together with a clear instruction and is asked to predict the SID of the next item the user might engage with.  
    \item \textbf{Alignment Tasks}: A collection of bridging tasks enforces a two-way mapping between natural language and SID space, grounding the discrete codes in text while injecting linguistic knowledge into their embeddings.  
\end{itemize}
Tasks from both groups are optimized jointly throughout the SFT stage and the subsequent RL stage. During RL, we adopt constrained decoding so that the model can only produce tokens from a predefined list containing every item’s SID and its canonical title. This constraint guarantees valid outputs and enables straightforward, rule-based reward computation. Detailed examples of the prompts are provided in the Appendix \ref{appendix:align_prompt}.

\subsection{Reinforced Preference Optimization} \label{sec:optimization}
After SFT, we further polish the policy with GRPO~\citep{deepseekmath,deepseek}. GRPO differs from classic RLHF in that it draws multiple candidates per prompt and normalizes rewards within the group, which reduces gradient variance.

Steps:
(1) For every prompt $x\!\sim\!D$, the frozen policy $\pi_{\theta_{\text{old}}}$ is rolled out $G$ times, yielding $\mathcal{Y}(x)=\{y^{(1)},\ldots,y^{(G)}\}$.  
(2) Each candidate $y^{(i)}$ is assigned a scalar score $S_i$.  
(3) Advantages are standardized inside the group:
\[
\hat{A}_i=\frac{S_i-\mu_{1:G}}{\sigma_{1:G}},
\]
where $\mu_{1:G}$ and $\sigma_{1:G}$ denote the mean and standard deviation of the $G$ rewards.

The surrogate objective becomes
\[
\small
\begin{aligned}
J_{\text{GRPO}}(\theta)=
\mathbb{E}_{x,y^{(i)}}
\biggl[
\frac{1}{G}\sum_{i=1}^{G}\frac{1}{|y^{(i)}|}
\sum_{t=1}^{|y^{(i)}|}
\!\Bigl(
\min\!\bigl(w_{i,t}\hat{A}_{i,t},\,
\operatorname{clip}(w_{i,t},1-\epsilon,1+\epsilon)\hat{A}_{i,t}\bigr)
-\beta\,\mathrm{KL}\!\bigl[\pi_\theta\|\pi_{\text{ref}}\bigr]
\Bigr)
\biggr],
\end{aligned}
\]
with token-level importance ratio  
$w_{i,t}= \dfrac{\pi_\theta(y^{(i)}_t\!\mid\!x,y^{(i)}_{<t})}
                 {\pi_{\theta_{\text{old}}}(y^{(i)}_t\!\mid\!x,y^{(i)}_{<t})}$.
The parameter $\epsilon$ controls clipping, while $\beta$ keeps the updated policy near a reference model through a KL term.

Applying RL with verifiable rewards (RLVR) to recommendation brings two obstacles:

\begin{itemize}
    \item \textbf{Unique generation space}. The action space is a closed set of item SIDs, orders of magnitude smaller than natural-language vocabularies. Re-sampling tends to produce duplicates, wasting computation. We therefore mix dynamic sampling \citep{DAPO} with constrained beam search to enlarge coverage while keeping outputs valid.
    \item \textbf{Sparse ranking supervision}. A hard binary reward (1 for the correct item, 0 otherwise) offers little guidance on ranking quality. We introduce an auxiliary ranking-shaped reward to penalize harder negatives with lower scores. Additionally, dense signals such as semantic similarity and collaborative scores are explored to furnish richer supervision.
\end{itemize}
\subsubsection{Sampling Strategy}
\label{subsection:sampling}

A practical obstacle when porting RLVR to recommendation is the poor sampling diversity brought by the limited action space: querying the policy multiple times with the same prompt often returns identical items, so the model observes few distinct negatives. We measure diversity via
\[
\mathrm{Div}\!\left(\{e_k\}_{k=1}^G\right)=
\frac{\bigl|\text{Unique}\bigl(\{e_k\}_{1}^{G}\bigr)\bigr|}{G},
\]
where the numerator counts unique items among the $G$ generations. Higher values indicate richer supervision.

Two complementary remedies are investigated:

\begin{itemize}
    \item \textbf{Dynamic Sampling} \citep{DAPO}. We first over-sample, then pick a subset that (i) must include the ground-truth item and (ii) maximizes internal diversity. Although helpful, this demands extra forward passes and still deteriorates as training progresses.

    \item \textbf{Beam Search}. We ultimately switch to beam search without length normalization \citep{D3, ReRe}. By construction, all beams differ, so the method guarantees zero duplication within each group and yields better diversity–efficiency trade-offs.
\end{itemize}

Based on our findings, MiniOneRec ultimately employs constrained beam search as its default sampler, ensuring that every generated item is valid while still providing a diverse set of candidate trajectories.

\subsubsection{Reward Design}
\label{method:reward}

Recommendation models are usually judged with ranking measures such as NDCG. In contrast, the standard GRPO setup supplies a binary reward, giving a value of $1$ to the true item and $0$ to every other candidate. This strategy treats all negatives as equally harmful. Earlier studies \citep{SGL,sdpo} have shown that focusing on hard negatives produces stronger rankers. Motivated by these results, we introduce a rank-aware reward that assigns different penalties based on how prominently a negative item appears in the model’s own ranking \citep{ReRe}.

Given a negative candidate $e_k$ whose generation probability ranks $\rho_k$ (with $\rho=1$ the most probable), we set
\[
\tilde{R}_{\text{rank}}(e_k,e_t)=
\begin{cases}
0, & e_k=e_t,\\
-\dfrac{1}{\log(\rho_k+1)}, & \text{otherwise},
\end{cases}\;\;
R_{\text{rank}}(e_k,e_t)=
-\dfrac{\tilde{R}_{\text{rank}}(e_k,e_t)}
{\sum_{j=1}^{G}\tilde{R}_{\text{rank}}(e_j,e_t)}.
\]
Thus, negatives that the model was very confident about (low $\rho_k$) receive stronger penalties.

The final reward combines rule-based and ranking components:
\[
R(e_k,e_t)=R_{\text{rule}}(e_k,e_t)+R_{\text{rank}}(e_k,e_t),
\]
where the rule-based term is
\[
R_{\text{rule}}(e_k,e_t)=
\begin{cases}
1, & e_k = e_t,\\
0, & \text{otherwise}.
\end{cases}
\]
The mixed reward discussed combines binary correctness with a soft ranking term, helping the LLM tell hard negatives apart. Yet recommendation data contain additional, under-utilized cues. To see whether such information can enhance or even replace the rule-based component in RLVR, we experiment with the following collaborative reward. Specifically, for every item suggested by the policy, we obtain its logit from a pre-trained collaborative-filtering model and pass that score back as reward, thereby injecting knowledge extracted from historical user–item interactions. MiniOneRec ultimately adopts the combined ranking-and-rule reward as its default choice.

\section{Training}
\label{training_detail}

MiniOneRec first converts item text into discrete codes. Titles and descriptions are embedded by the Qwen3-Embedding-4B encoder, after which RQ-VAE performs residual quantization. This tokenizer is trained on a single GPU with a batch size of 20\,480, a learning rate of $1\times10^{-3}$, and 10,000 training epochs. The resulting SID vocabulary is plugged into a Qwen2.5-Instruct backbone. SFT takes place on eight NVIDIA H100, each holding 128 samples. Training lasts up to ten epochs with early stopping (patience one epoch). The initial learning rate is $3\times10^{-4}$ and follows a cosine decay schedule. Starting from the SFT checkpoint, we apply GRPO for two additional epochs while keeping the KL weight $\beta$ unchanged. During roll-out, beam search with width 16 is adopted, so every input generates sixteen distinct candidate sequences.

In the performance comparison with existing recommenders, conventional recommenders are trained with binary cross-entropy and the Adam optimizer; learning rates are chosen from \{1e-2, 1e-3, 1e-4\}, and the weight-decay term is scanned over \{1e-2, 1e-3, 1e-4, 1e-5, 1e-6\}. The batch size is 1\,024.  
For TIGER, we rely on a T5 encoder–decoder and reuse Qwen3-Embedding-4B for item embeddings.  
All LLM-powered systems, including ours, share the Qwen2.5-Instruct backbone and use AdamW. Batches contain 128 examples for SFT and preference alignment, and 512 samples for RL. The learning rate is $3\times10^{-4}$ during SFT, while S-DPO and RL use $1\times10^{-5}$. S-DPO runs for one epoch with $\beta=0.1$ and samples three negative items; D$^{3}$ tries interpolation factors $\alpha$ of 0.8, 0.9, and 1.0.

\section{Evaluation}
\label{results}

This section details our empirical study. We begin with a scaling investigation on two real-world benchmarks \citep{amazon}, examining how MiniOneRec’s loss curves change as model size grows. We then benchmark MiniOneRec against three groups of baselines: traditional sequential recommenders, SID-based generators, and recent LLM-driven systems.  
To evaluate the model’s capability for cross-domain recommendation generalization, we frame SID next-item prediction as a task of recommendation rule discovery and evaluate the model on domains that never appeared during training.  
Comprehensive ablation tests follow, which help us locate the parts of the framework that contribute most to the final score. Finally, we look into how the broad knowledge stored in pre-trained LLM weights affects generative recommendation.
In summary, we study the following questions within this section:

\begin{itemize}
    \item How does the loss of MiniOneRec scale with different model size?
    
    \item How does MiniOneRec perform in comparison to other baseline methods?
    
    \item How does MiniOneRec perform under completely unseen domains?
    
    \item How do the designed components contribute to MiniOneRec's recommendation efficiency?

    \item How do the pre-trained weights of the LLM influence the performance of generative recommendation?

\end{itemize}

\subsection{Evaluation Setup}


Experiments are carried out on two real-world slices of the Amazon Review dataset \citep{amazon}, namely \textit{Office} and \textit{Industrial}. To measure top-$K$ recommendation accuracy, we follow standard practice and compute Hit Rate (HR@K) as well as Normalized Discounted Cumulative Gain (NDCG@K).

\begin{table}[t]
\caption{Performance of MiniOneRec Compared to Traditional Methods, Generative Methods, and LLM-based Methods} 
\label{tab:overall}
\centering
\begin{adjustbox}{max width=0.9\textwidth}
\renewcommand{\arraystretch}{1.00}
\begin{tabular}{cccccccc}
\toprule
\multicolumn{1}{c}{Datasets}            & Methods    & HR@3   & NDCG@3 & HR@5   & NDCG@5 & HR@10  & NDCG@10 \\ \midrule
\multirow{14}{*}{\textbf{Industrial}} & \multicolumn{7}{c}{\textbf{Traditional}}                          \\ \cmidrule(lr){2-8} 
                                      & GRU4Rec    & 0.0638 & 0.0542 & 0.0774 & 0.0598 & 0.0999 & 0.0669  \\
                                      & Caser      & 0.0618 & 0.0514 & 0.0717 & 0.0555 & 0.0942 & 0.0628  \\
                                      & SASRec     & 0.0790 & 0.0700 & 0.0909 & 0.0748 & 0.1088 & 0.0806  \\ \cmidrule(lr){2-8}
                                      & \multicolumn{7}{c}{\textbf{Generative}}                           \\ \cmidrule(lr){2-8}
                                      & HSTU       & 0.0927 & 0.0885 & 0.1037 & 0.0918 & 0.1163 & 0.0958  \\
                                      & TIGER      & 0.0852 & 0.0742 & 0.1010 & 0.0807 & 0.1321 & 0.0908  \\
                                      & LCRec      & 0.0915 & 0.0805 & 0.1057 & 0.0862 & 0.1332 & 0.0952  \\ \cmidrule(lr){2-8}
                                      & \multicolumn{7}{c}{\textbf{LLM-based}}                            \\ \cmidrule(lr){2-8}
                                      & BIGRec     & 0.0931 & 0.0841 & 0.1092 & 0.0907 & 0.1370 & 0.0997  \\
                                      & D$^3$         & 0.1024 & \uline{0.0991} & 0.1213 & 0.0989 & 0.1500 & \uline{0.1082} \\
                                      & S-DPO      & \uline{0.1032} & 0.0906 & \uline{0.1238} & \uline{0.0991} & \uline{0.1524} & \uline{0.1082}  \\ \cmidrule(lr){2-8}
                                      & \multicolumn{7}{c}{\textbf{Ours}}                                 \\ \cmidrule(lr){2-8}
                                      & \cellcolor[HTML]{ECF4FF}\textbf{MiniOneRec}      & \cellcolor[HTML]{ECF4FF}\textbf{0.1143} & \cellcolor[HTML]{ECF4FF}\textbf{0.1011} & \cellcolor[HTML]{ECF4FF}\textbf{0.1321} & \cellcolor[HTML]{ECF4FF}\textbf{0.1084} & \cellcolor[HTML]{ECF4FF}\textbf{0.1586} & \cellcolor[HTML]{ECF4FF}\textbf{0.1167}  \\
                                      \midrule
\multirow{14}{*}{\textbf{Office}}     & \multicolumn{7}{c}{\textbf{Traditional}}                          \\ \cmidrule(lr){2-8}
                                      & GRU4Rec    & 0.0629 & 0.0528 & 0.0789 & 0.0595 & 0.1019 & 0.0669  \\
                                      & Caser      & 0.0748 & 0.0615 & 0.0865 & 0.0664 & 0.1093 & 0.0737  \\
                                      & SASRec     & 0.0861 & 0.0769 & 0.0949 & 0.0805 & 0.1120 & 0.0858  \\ \cmidrule(lr){2-8}
                                      & \multicolumn{7}{c}{\textbf{Generative}}                           \\ \cmidrule(lr){2-8}
                                      & HSTU       & 0.1134 & 0.1031 & 0.1252 & 0.1079 & 0.1400 & 0.1126  \\
                                      & TIGER      & 0.0986 & 0.0852 & 0.1163 & 0.0960 & 0.1408 & 0.1002  \\
                                      & LCRec      & 0.0921 & 0.0807 &    0.1048 & 0.0859 & 0.1237 & 0.0920  \\ \cmidrule(lr){2-8}
                                      & \multicolumn{7}{c}{\textbf{LLM-based}}                            \\ \cmidrule(lr){2-8}
                                      & BIGRec     & 0.1069 & 0.0961 & 0.1204 & 0.1017 & 0.1434 & 0.1091  \\
                                      & D$^{3}$         & \uline{0.1204} & \uline{0.1055} & \uline{0.1406} & \uline{0.1139} & \textbf{0.1634} & 0.1213  \\
                                      & S-DPO      & 0.1169 & 0.1033 & 0.1356 & 0.1110 & \uline{0.1587} & \textbf{0.1255}  \\ \cmidrule(lr){2-8}
                                      & \multicolumn{7}{c}{\textbf{Ours}}                                 \\ \cmidrule(lr){2-8}
                                      & \cellcolor[HTML]{ECF4FF}\textbf{MiniOneRec}      & \cellcolor[HTML]{ECF4FF}\textbf{0.1217} & \cellcolor[HTML]{ECF4FF}\textbf{0.1088} & \cellcolor[HTML]{ECF4FF}\textbf{0.1420} & \cellcolor[HTML]{ECF4FF}\textbf{0.1172} & \cellcolor[HTML]{ECF4FF}\textbf{0.1634} & \cellcolor[HTML]{ECF4FF}\uline{0.1242}  \\
                                      \bottomrule
\end{tabular}
\end{adjustbox}
\end{table}

\subsection{Scaling}
\begin{wrapfigure}[14]{r}{0.48\columnwidth}   
    \centering
    \vspace{2pt}                             
    \includegraphics[width=\linewidth]{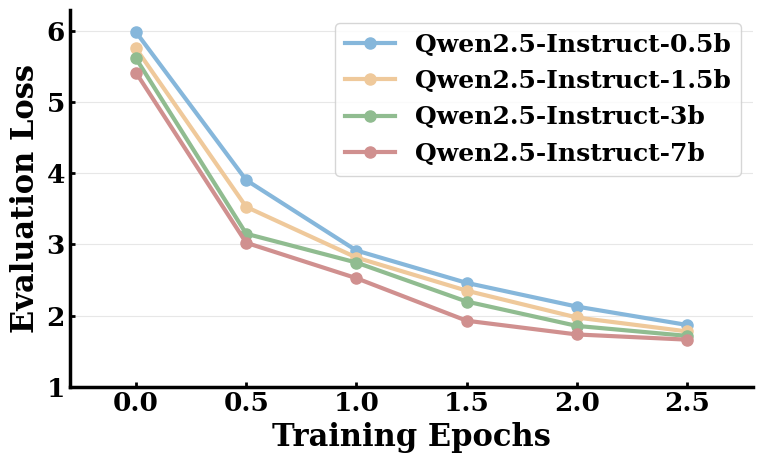}
    \vspace{-16pt}                             
    \caption{Evaluation loss vs.\ SFT training epoch}
    \label{fig:scaling_eval_loss}
\end{wrapfigure}

We demonstrate the scaling capabilities of MiniOneRec, where the convergence loss decrease consistently as the model size increases. 
As Figure \ref{fig:scaling} shows, under the generative recommendation paradigm, there is a clear correlation between model size and loss. As the LLM scale increases, the convergence loss continues to decrease. Furthermore, Figure \ref{fig:scaling_eval_loss} tracks the evaluation loss on the SFT training set, recorded every 0.5 epoch. It is evident that models with larger parameter counts maintain lower evaluation losses throughout nearly the entire training process and converge more rapidly.
Such a superior scaling effect demonstrates the potential of generative recommenders as the next-generation recommendation models.

\subsection{Performance Comparison}



Our baselines contain three categories: (1) Traditional recommendation models, including GRU4Rec \citep{GRU4Rec}, Caser \citep{Caser}, SASRec \citep{SASRec}; (2) Generative recommendation models: HSTU \citep{HSTU}, TIGER \citep{TIGER}, LC-Rec \citep{LCRec}; (3) LLM-based recommendation models, including BigRec \citep{BIGRec}, D$^3$ \citep{D3}, S-DPO \citep{sdpo}.

We evaluate \textsc{MiniOneRec} on the \textit{Industrial} and \textit{Office} benchmarks and summarize the outcomes in Table~\ref{tab:overall}. Two key insights stand out:

\begin{itemize}
    \item \textbf{Utility of LLM World Knowledge.} Recommenders powered by LLMs, such as BIGRec and D$^{3}$, clearly surpass traditional systems like GRU4Rec and Caser, showing that the broad knowledge embedded in LLMs translates into better recommendation accuracy.  
    \item \textbf{Effectiveness of MiniOneRec.} MiniOneRec goes a step further: by aligning the full generation process with the task objective and using reinforced preference optimization during RL, it consistently outperforms previous generative solutions across most reported metrics. Moreover, by operating in the compact SID space rather than on verbose textual titles, MiniOneRec requires substantially fewer context tokens and yields faster inference, translating into lower latency and smaller memory footprints at serving time.
\end{itemize}

\subsection{Transferability}
\label{subsec: Transferability}

We evaluate the out-of-distribution (OOD) robustness of MiniOneRec through an experiment we call \textit{SID pattern discovery}. The model is trained exclusively on the \textit{Industrial} domain and then deployed, without any further tuning, to the never-seen \textit{Office} domain. Prior work \citep{RLvsSFT, RLReallyLearn, SFT-then-RL, RevisitingRL} suggests that SFT may overfit to the source domain and hurt transfer, so we add an RL-only variant, MiniOneRec-w/ RL-OOD, that skips SFT entirely to emphasize generalization.
The comparison involves four systems:  
(1) GRU4Rec, trained and tested inside the \textit{Office} domain;  
(2) Qwen-Text, which represents user histories as plain sentences and predicts the next item by its title with no additional fine-tuning;  
(3) Qwen-SID, which encodes the same history with SID tokens and produces the next SID with no additional fine-tuning;  
(4) MiniOneRec-w/ RL-OOD, trained via GRPO on \textit{Industrial} only and evaluated on \textit{Office}, to evaluate its OOD (out-of-distribution) performance.

Table~\ref{tab:ood} reports the outcomes. Qwen-Text performs poorly, while Qwen-SID does noticeably better, indicating that a structured SID vocabulary is easier for an LLM to exploit. Although MiniOneRec-w/ RL-OOD falls short of the full MiniOneRec on in-domain scores, its reinforcement-only training grants excellent transfer, achieving competitive accuracy on the unseen catalog. Despite the substantial domain shift and possible semantic drift among SIDs, MiniOneRec successfully uncovers reusable interaction patterns, underscoring the framework’s promise for cross-domain recommendation.



\subsection{Ablation Study}

To validate the effectiveness of each component in the MiniOneRec framework, we compare it with the following alternative approaches. 

\subsubsection{Aligning Strategy}

\begin{table}[]
\caption{Performance of MiniOneRec and its variants on completely unseen recommendation domains} 
\label{tab:ood}
\resizebox{\textwidth}{!}{
\begin{tabular}{cccccccc}
\toprule
Dataset                           & Method                          & HR@3                           & NDCG@3                         & HR@5                           & NDCG@5                         & HR@10                          & NDCG@10                        \\ \hline
                                  & \cellcolor[HTML]{EFEFEF}GRU4Rec & \cellcolor[HTML]{EFEFEF}0.0629 & \cellcolor[HTML]{EFEFEF}0.0528 & \cellcolor[HTML]{EFEFEF}0.0789 & \cellcolor[HTML]{EFEFEF}0.0595 & \cellcolor[HTML]{EFEFEF}0.1019 & \cellcolor[HTML]{EFEFEF}0.0669 \\
                                  & Qwen-Text                       & 0.0031                         & 0.0021                         & 0.0044                         & 0.0026                         & 0.0057                         & 0.0030                         \\
                                  & Qwen-SID                        & 0.0300                         & 0.0214                         & 0.0456                         & 0.0282                         & 0.0733                         & 0.0373                         \\
\multirow{-4}{*}{\textbf{Office}} & \cellcolor[HTML]{ECF4FF}MiniOneRec-w/ RL-OOD              & \cellcolor[HTML]{ECF4FF}0.0553                         & \cellcolor[HTML]{ECF4FF}0.0433                         & \cellcolor[HTML]{ECF4FF}0.0691                         & \cellcolor[HTML]{ECF4FF}0.0489                         & \cellcolor[HTML]{ECF4FF}0.0892                         & \cellcolor[HTML]{ECF4FF}0.0553                         \\ \hline
\end{tabular}
}
\end{table}

To isolate the impact of each component, we compare the full MiniOneRec with three pared-down variants.  
(1) \textsc{MiniOneRec}--\textsc{w/o~Align}: removes any language–SID alignment and treats recommendation purely as a SID-to-SID task.  
(2) \textsc{MiniOneRec}--\textsc{w/~SFTAlign}: keeps the alignment objective during SFT stage only, while RL uses SID data alone.  
(3) \textsc{MiniOneRec}--\textsc{w/~RLAlign}: SFT relies solely on SID supervision, and the alignment tasks are introduced later in the RL stage.

Figure~\ref{fig:ab1} summarizes the findings. The complete MiniOneRec, which maintains alignment throughout the whole pipeline, delivers the highest scores on every metric. The \textsc{MiniOneRec}--\textsc{w/o~Align} performs worst, indicating that grounding SID generation in world knowledge is essential.

\subsubsection{Sampling Strategy}

We study how different roll-out methods affect MiniOneRec by switching only the trajectory generator while keeping everything else fixed:  
(1) \textsc{MiniOneRec}--\textsc{Common} relies on a plain Top-$k$ decoder to produce exactly the required number of paths.  
(2) \textsc{MiniOneRec}--\textsc{Dynamic} follows our two-step sampler: it first draws one and a half times the budget, then retains as many unique items as possible for RL.  
The full model adopts beam search with width 16.

As plotted in Figure~\ref{fig:ab2}, the complete MiniOneRec delivers the highest accuracy while using roughly two-thirds of the samples needed by the dynamic variant, demonstrating that beam search is the most cost-efficient choice among the tested strategies.

\subsubsection{Reward Design}

Three variants are compared:
(1) \textsc{MiniOneRec}--\textsc{w/~ACC} that relies solely on a binary correctness signal;  
(2) \textsc{MiniOneRec}--\textsc{w/~Collaborative} that replaces the ranking term with logits taken from a frozen SASRec model so as to supply collaborative cues;  

\begin{figure*}[t]
  \centering
  \begin{subfigure}[t]{0.28\textwidth}
    \centering
    \includegraphics[width=\textwidth]{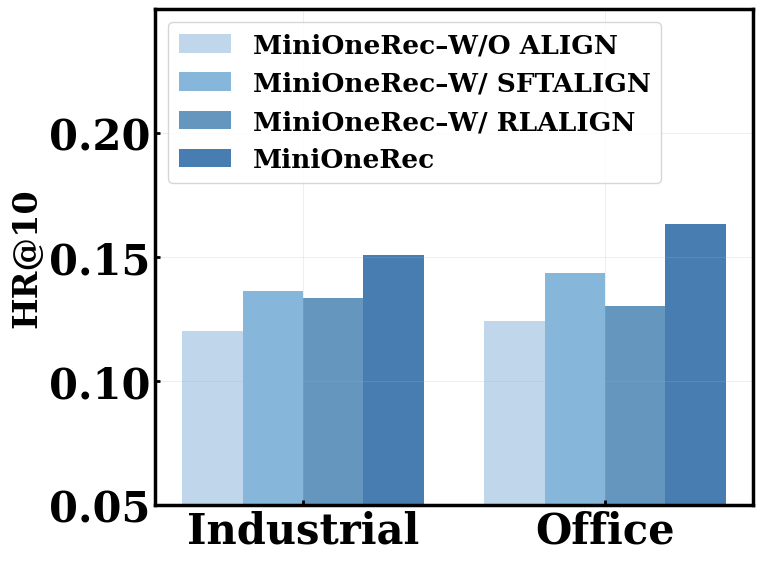}
    \caption{Aligning Strategy.}
    \label{fig:ab1}
\end{subfigure}\hspace{2mm}
\begin{subfigure}[t]{0.28\textwidth}
    \centering
        \includegraphics[width=\textwidth]{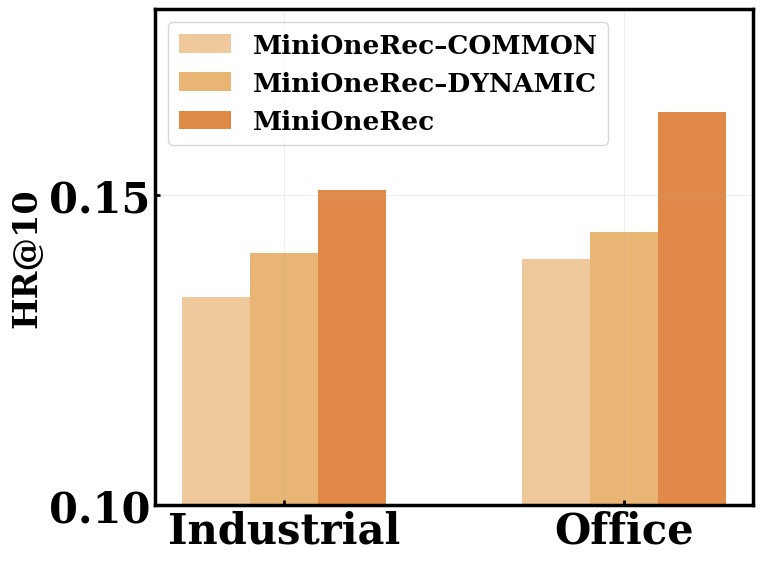}
    \caption{Sampling Strategy.}
    \label{fig:ab2}
\end{subfigure}
\centering
  \begin{subfigure}[t]{0.28\textwidth}
    \centering
    \includegraphics[width=\textwidth]{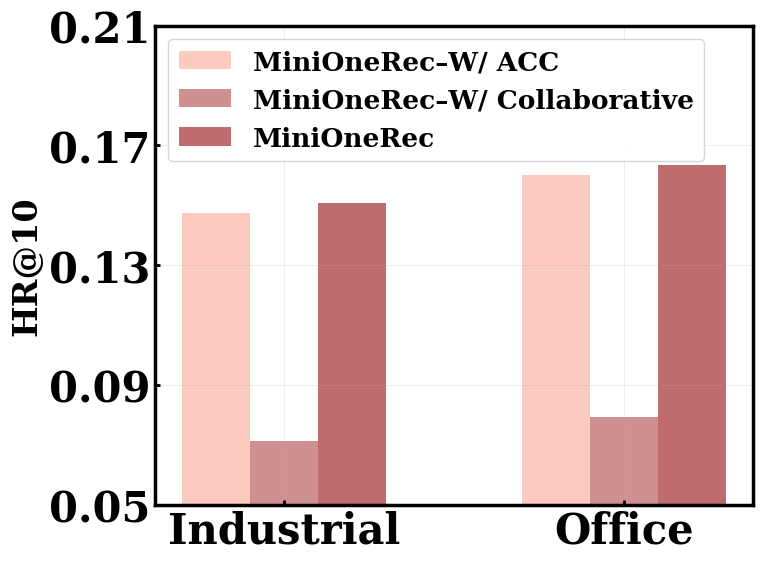}
    \caption{Reward Design.}
    \label{fig:ab3}
\end{subfigure}\hspace{2mm}
  \caption{Study on the effectiveness of MiniOneRec’s individual components.
Figure \ref{fig:ab1} examines model performance under different alignment strategies;
Figure \ref{fig:ab2} investigates various sampling strategies;
Figure \ref{fig:ab3} evaluates the impact of alternative reward designs.}
  \label{fig:ablation}
    \vspace{-10pt}
\end{figure*}

As shown in the Figure \ref{fig:ab3}, the full MiniOneRec achieves the best overall performance. To be noticed, injecting the collaborative reward information into the RL process instead led to significant degradation. We hypothesize that, this stems from reward hacking: as recommendation accuracy declines, the reward continues to increase, revealing a misalignment between this collaborative reward signal and the true objective.

\begin{table}[]
\caption{Performance of MiniOneRec initialized with pre-trained weights versus random initialization across two domains} 
\label{tab:pretrain-impact}
\resizebox{\textwidth}{!}{
\begin{tabular}{clllllll}
\toprule
\multicolumn{1}{l}{Datesets}          & Methods            & HR@3            & NDCG@3          & HR@5            & NDCG@5          & HR@10           & NDCG@10         \\ \midrule
\multirow{2}{*}{\textbf{Industrial}} & MiniOneRec-scratch & 0.0757          & 0.0672          & 0.0891          & 0.0726          & 0.1134          & 0.0804          \\
                                     & \cellcolor[HTML]{ECF4FF}\textbf{MiniOneRec}         & \cellcolor[HTML]{ECF4FF}\textbf{0.1125} & \cellcolor[HTML]{ECF4FF}\textbf{0.0988} & \cellcolor[HTML]{ECF4FF}\textbf{0.1259} & \cellcolor[HTML]{ECF4FF}\textbf{0.1046} & \cellcolor[HTML]{ECF4FF}\textbf{0.1546} & \cellcolor[HTML]{ECF4FF}\textbf{0.1139} \\ \midrule
\multirow{2}{*}{\textbf{Office}}     & MiniOneRec-scratch & 0.0959          & 0.0855          & 0.1057          & 0.0896          & 0.1196          & 0.0941          \\
                                     & \cellcolor[HTML]{ECF4FF}\textbf{MiniOneRec}         & \cellcolor[HTML]{ECF4FF}\textbf{0.1217} & \cellcolor[HTML]{ECF4FF}\textbf{0.1088} & \cellcolor[HTML]{ECF4FF}\textbf{0.1420} & \cellcolor[HTML]{ECF4FF}\textbf{0.1172} & \cellcolor[HTML]{ECF4FF}\textbf{0.1634} & \cellcolor[HTML]{ECF4FF}\textbf{0.1242} \\ \bottomrule
\end{tabular}
}
\end{table}

\subsection{Pre-trained LLM Impact}
\label{llmmatter}
To isolate the impact of pre-trained LLMs on generative recommendation, we instantiate two variants of MiniOneRec: one initialized from a general-purpose pre-trained LLM and the other trained from scratch with random weights. As shown in Table \ref{tab:pretrain-impact}, Experiments on two benchmark datasets reveal a consistent pattern: the model that starts from pre-trained weights significantly outperforms its randomly initialized counterpart. We conjecture that (i) the general reasoning ability acquired during large-scale language pre-training allows the model to cast the next-SID prediction task as a problem of pattern discovery, as discussed in Section~\ref{subsec: Transferability}, and (ii) the factual knowledge already encoded in the LLM offers a head start in understanding the real-world semantics behind each SID, part of which can be transferred to the recommendation domain.
\section{Conclusion}
\label{conclusion}
This report introduces \textbf{MiniOneRec}, the first fully open-source generative recommendation framework, which to the best of our knowledge provides an end-to-end workflow spanning SID construction, SFT, and recommendation-oriented RL.
By systematically validating scaling laws on public benchmarks, we demonstrate that larger generative recommenders achieve lower training and evaluation losses than their smaller counterparts, confirming the parameter–efficiency advantage of the SID-based paradigm over traditional embedding-centric models.  
Building on this insight, we introduce post-training techniques:  
(i) full-process SID alignment, which embeds SID tokens into the model vocabulary and imposes auxiliary alignment tasks across both SFT and RL stages, and (ii) reinforced preference optimization, which combines constrained decoding, beam-based sampling, and hybrid reward design.  
Extensive experiments on Amazon Review show that MiniOneRec consistently surpasses strong sequential, generative, and LLM-based baselines, while maintaining a lean post-training footprint.

Looking forward, we will keep maintaining and extending the MiniOneRec codebase.  
A public roadmap will guide future developments, and we warmly welcome community contributions.  
Planned updates include new datasets, more advanced tokenisation schemes, larger backbone models, and enhanced training pipelines, ensuring that MiniOneRec remains a solid reference platform for research and practice in large-scale generative recommendation.



\newpage
\bibliographystyle{unsrtnat}
\bibliography{iclr2026_conference}

\newpage
\appendix

\section{Alignment Prompts}
\label{appendix:align_prompt}

\subsection{Overview}
In this section we describe how we \textbf{align the world knowledge of a LLM with the SID space} so that the model can be directly used for generative recommendation.  

We append a three–layer codebook, each layer containing $256$ unique SIDs, to the original tokenizer vocabulary. The inserted codes are treated as indivisible tokens, enabling the LLM to read or write SID sequences without any sub-word splitting.
Then we introduce a set of auxiliary \emph{Alignment Tasks} to bridge the semantic gap between the textual vocabulary and the newly added SIDs. 
A specialized \emph{Recommendation Tasks} further teach the model to predict the SID of the next item given an user history.

\subsection{Recommendation Tasks}

\begin{enumerate}
    \item \textbf{Generative Retrieval.} The main task ofor generative recommendation.  
    The LLM receives a chronologically ordered SID sequence that represents the user’s recent interactions, together with an explicit instruction such as “Recommend the next item.”, and is asked to predict the SID of the next item the user might engage with. A sample prompt is shown in the Figure \ref{fig:sid2sid}.

\begin{figure}[H]
    \centering
    \includegraphics[width=0.8\textwidth]{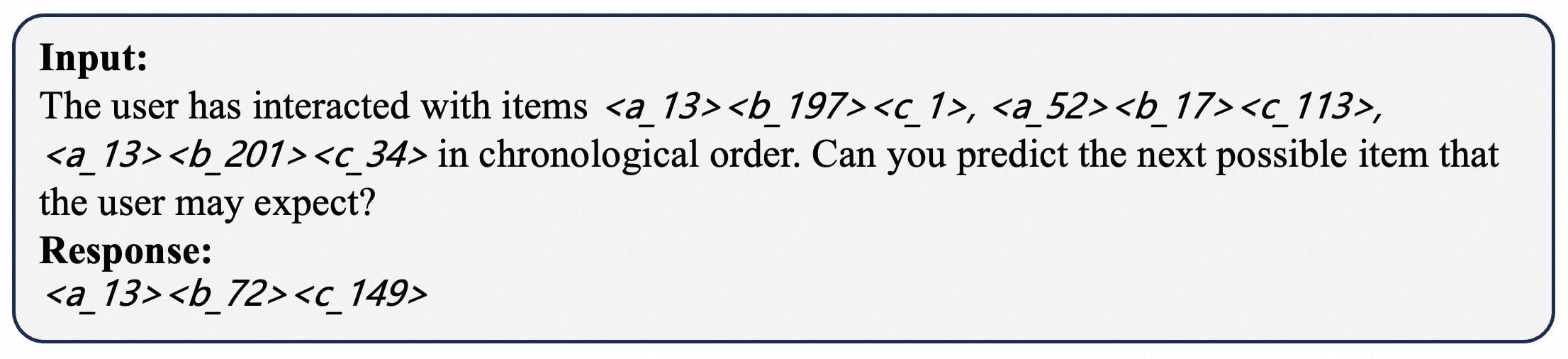}
    \vspace{-10pt}
    \caption{Generative Retrieval Prompt.}
    \label{fig:sid2sid}
    \vspace{-10pt}
\end{figure}

    \item \textbf{Asymmetric Item Prediction.}
    (a) Given a textual user history, predict the SID of the next item; (b) given a SID-only history, generate the textual title of the next item. Sample prompts are shown in the Figure \ref{fig:titlehis2sid} and Figure \ref{fig:sidhis2title}.

\begin{figure}[H]
    \centering
    \includegraphics[width=0.8\textwidth]{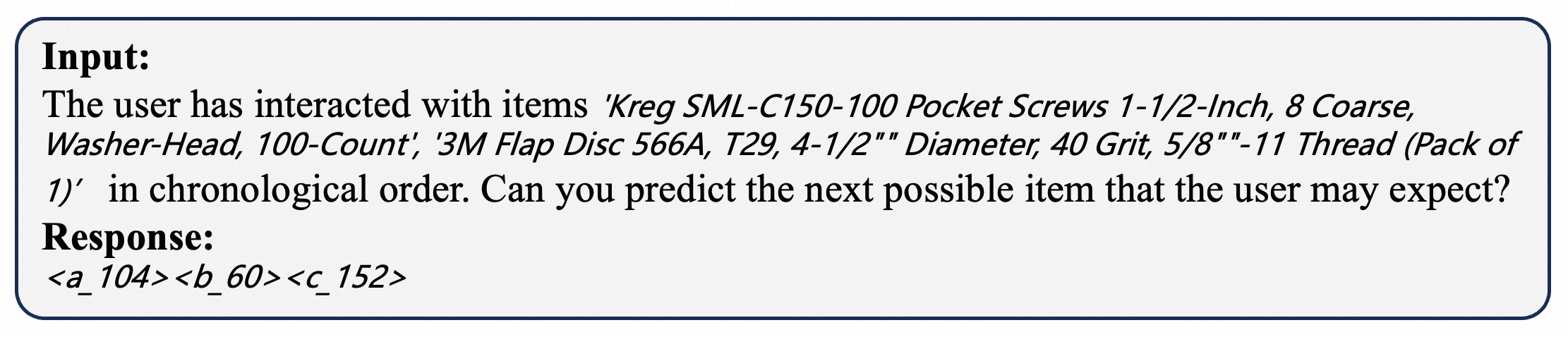}
    \vspace{-10pt}
    \caption{Asymmetric Item Prediction Prompt1.}
    \label{fig:titlehis2sid}
    \vspace{-10pt}
\end{figure}

\begin{figure}[H]
    \centering
    \includegraphics[width=0.8\textwidth]{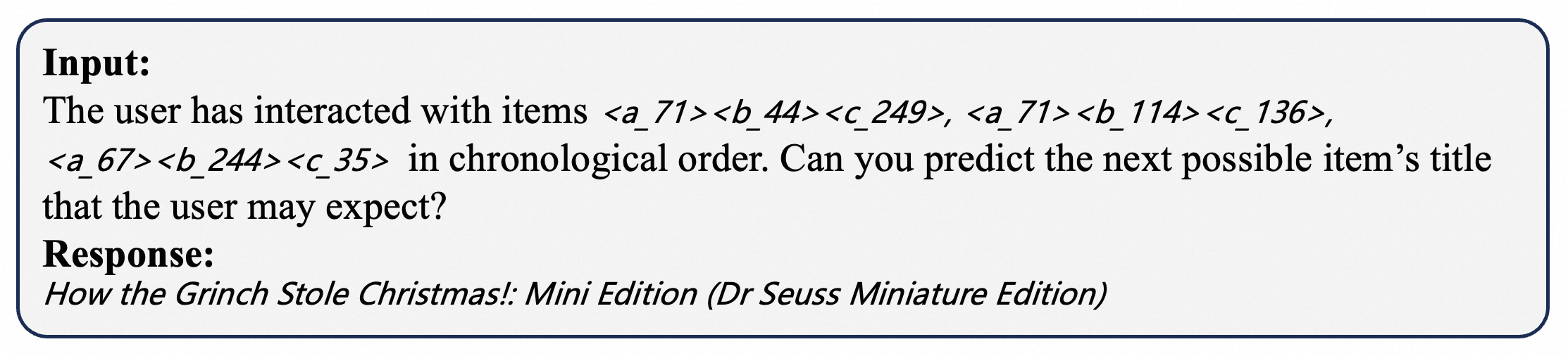}
    \vspace{-10pt}
    \caption{Asymmetric Item Prediction Prompt2.}
    \label{fig:sidhis2title}
    \vspace{-10pt}
\end{figure}
\end{enumerate}

\subsection{Alignment Tasks}

\begin{enumerate}
    \item \textbf{SID–Text Semantic Alignment.} (a) Predict an item's textual title from its SID; (b) predict the SID from the item's textual title.  Sample prompts are shown in the Figure \ref{fig:sid2title} and Figure \ref{fig:title2sid}.

\begin{figure}[H]
    \centering
    \includegraphics[width=0.45\linewidth]{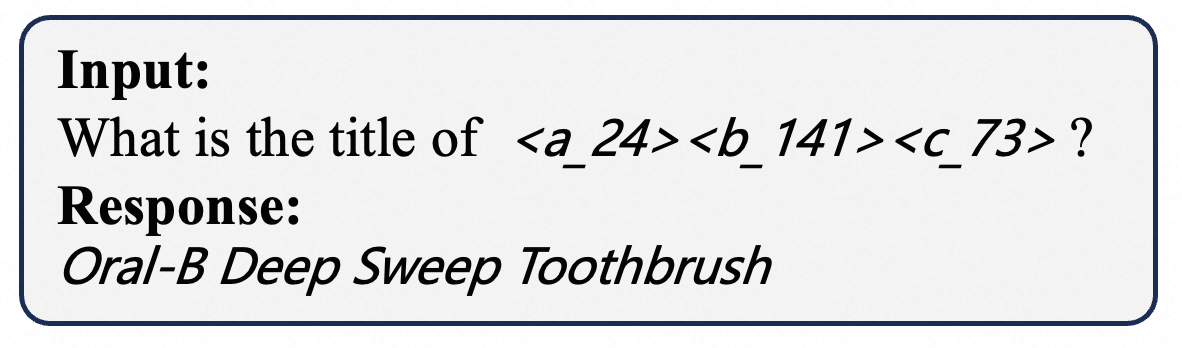}
    \vspace{-10pt}
    \caption{SID–Text Semantic Alignment Prompt1.}
    \label{fig:sid2title}
    \vspace{-10pt}
\end{figure}

\begin{figure}[H]
    \centering
    \includegraphics[width=0.64\textwidth]{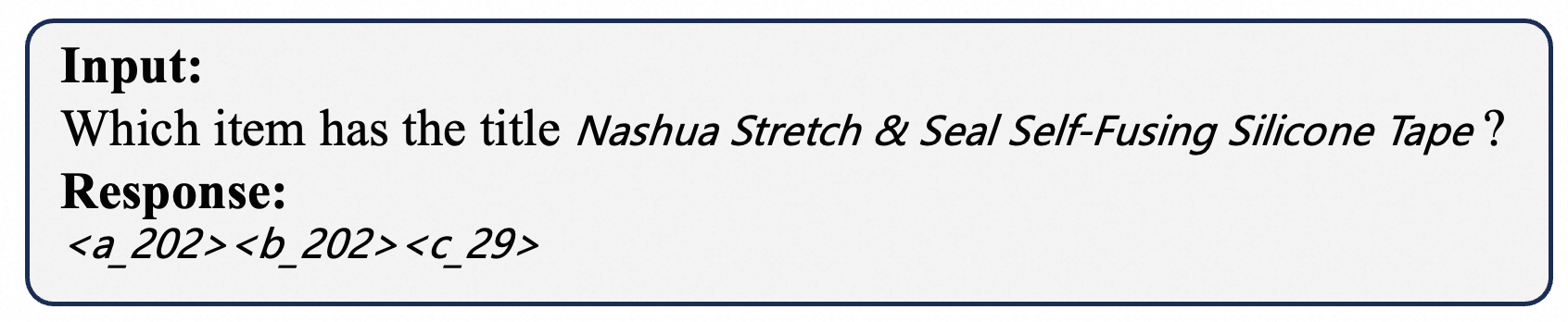}
    \vspace{-10pt}
    \caption{SID–Text Semantic Alignment Prompt2.}
    \label{fig:title2sid}
    \vspace{-10pt}
\end{figure}

    \item \textbf{Item Description Reconstruction.}  
          To ground SIDs in richer semantics, we ask the model to generate the item description from a single SID and, conversely, infer the SID from the description. We perform this task only during the SFT stage because the description space is large and diverse. Sample prompts is shown in the Figure \ref{fig:description2sid}.

\begin{figure}[H]
    \centering
    \includegraphics[width=0.8\textwidth]{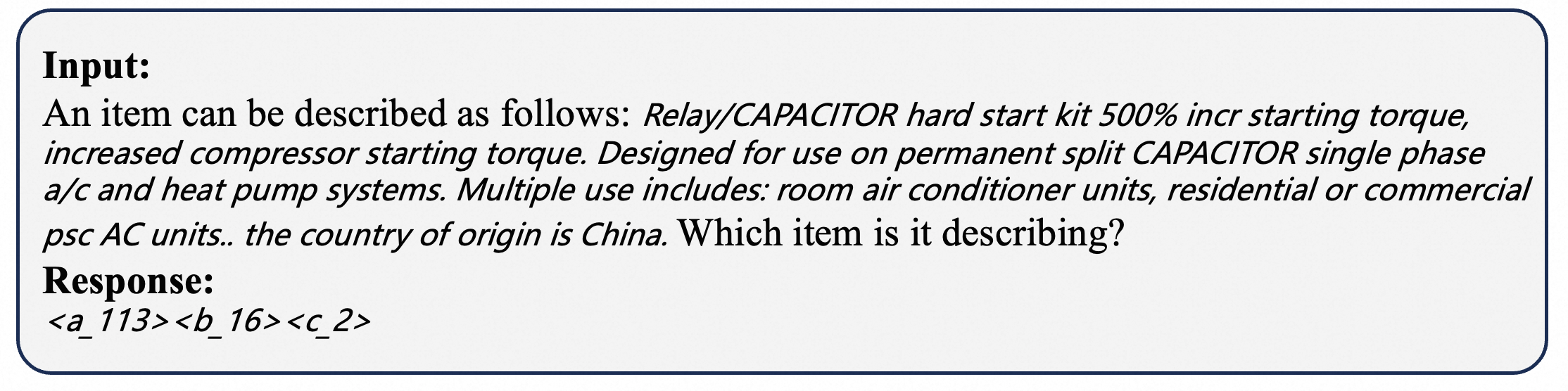}
    \vspace{-10pt}
    \caption{Item Description Reconstruction Prompt.}
    \label{fig:description2sid}
    \vspace{-10pt}
\end{figure}

    \item \textbf{User Preference Summarization.}  
          Given a sequence of SIDs, the model produces a short natural–language profile that summarizes the user’s interests. Because the raw dataset lacks explicit preference labels, we employ \textsc{DeepSeek} \citep{deepseek} to extract summaries from the item's meta data and users’ textual reviews and use them as pseudo labels. This task, too, is restricted to the SFT stage due to the open-ended output space.
\begin{figure}[H]
    \centering
    \includegraphics[width=0.8\textwidth]{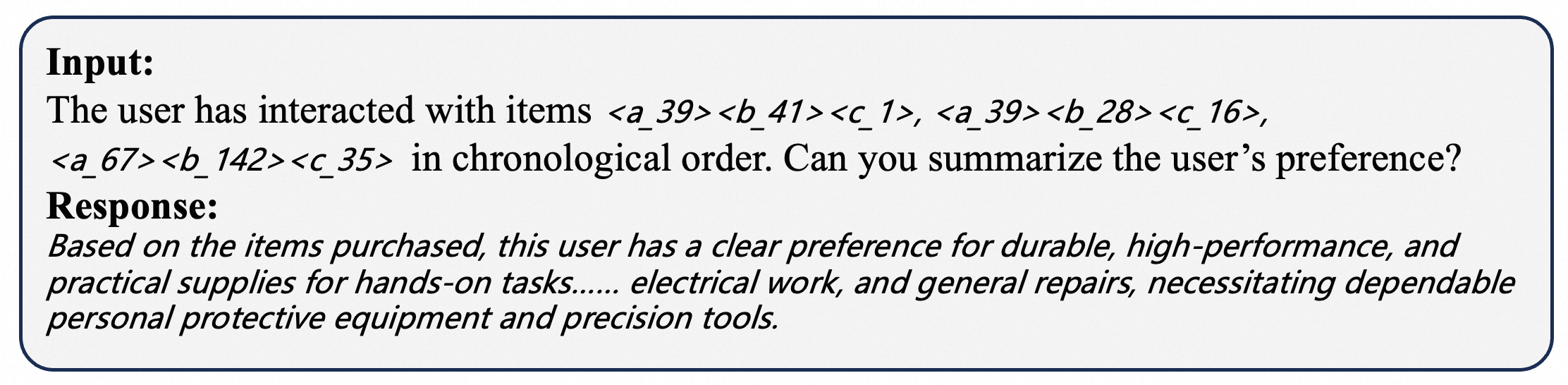}
    \vspace{-10pt}
    \caption{User Preference Summarization Prompt.}
    \label{fig:sid2title}
    \vspace{-10pt}
\end{figure}

\end{enumerate}

\section{Datasets}

We evaluate our model on two Amazon Review subsets:  Amazon Review subsets (\textit{Industrial\_and\_Scientific} and \textit{Office\_Products}).  
To keep computational costs affordable, we follow the trimming strategy used in \citep{D3}.  
The main steps are:  
(1) remove users and items with fewer than five interactions;  
(2) for \textit{Toys\_and\_Games}, keep events from October 2016 to November 2018;  
(3) for \textit{Industrial\_and\_Scientific}, which is smaller, keep all events between October 1996 and November 2018;   
(4) truncate every user history to at most ten items;  
(5) finally, split each dataset chronologically into training, validation and test sets with an 8:1:1 ratio.  
Key statistics of the resulting training splits are reported in Table~\ref{tab:stats}.

\begin{table}[H]
\centering
\caption{Statistics of datasets.} 
\label{tab:stats}
\begin{tabular}{ccc}
\toprule
Datasets & Inductrial & Office \\ \midrule
Items    & 3,685      & 3,459  \\
Train    & 3,6259     & 3,8924 \\
Valid    & 4,532      & 4,866  \\
Test     & 4,533      & 4,866  \\ \bottomrule
\end{tabular}
\end{table}

\end{document}